# Unit Cell Design for Aperiodic Metasurfaces

Jordan Budhu, *Member, IEEE*, Nicholas Ventresca, and Anthony Grbic, *Fellow, IEEE*

*Abstract*—A technique is presented for the design of printed unit cells in aperiodic metasurface environments. The method begins with a solved matrix equation governing electromagnetic scattering from a homogenized metasurface design. The matrix equation is used to find the local, inhomogeneous electric field exciting a printed-circuit unit cell geometry. The local field is then impressed onto the printed circuit geometry and the induced surface current numerically computed. The computed surface current is sampled at the matrix equation discretization. The matrix equation is then used to compute the electric field scattered by the printed-circuit unit cell onto its neighbors using the sampled current in place of the current of the original homogenized unit cell. The printed circuit geometry is optimized to scatter the same field as the homogenized unit cell when excited with the local electric field computed. Two design examples are provided. Both a finite-sized, wide-angle reflecting metasurface, and a metasurface reflectarray designed to scan and collimate an incident cylindrical wave, are realized with printed-circuit unit cells using the proposed approach. It is shown that the local periodicity approximation can be used to accurately design the unit cells of either finite-sized metasurface.

*Index Terms*—Metasurface, Aperiodic, Inhomogeneous, Unit Cell Design

## I. INTRODUCTION

SUBWAVELENGTH structured composite materials, also known as metamaterials, promise properties beyond what is normally found in nature, such as dielectric materials with negative refractive indices, materials exhibiting diamagnetism and paramagnetism at optical frequencies, gigantic optical activity, exceptionally large nonlinear optical susceptibilities, or non-reciprocal behavior [1]. Since losses accrue through field propagation in metamaterials, metasurfaces which can perform wavefront transformations over a subwavelength distance (effectively a surface) were proposed [2]. Constitutive surface parameters which perform these field transformations called Generalized Sheet Transition Conditions (GSTCs), offer a way to model homogenized metasurfaces [3], [4]. Homogenization refers to averaging the response of a unit cell (which in general contains a complex geometry) and assigning simple proportionality constants (impedances, admittances, and coupling coefficients) between induced currents and averaged fields. This way the metasurface can be described, analyzed, and designed in a reduced dimensional space. After a design is completed in the reduced dimensional space, the proportionality constants must be translated back to patterned/textured geometries by way of an extraction method. That is, the extraction method assigns proportionality constants to complex geometries.

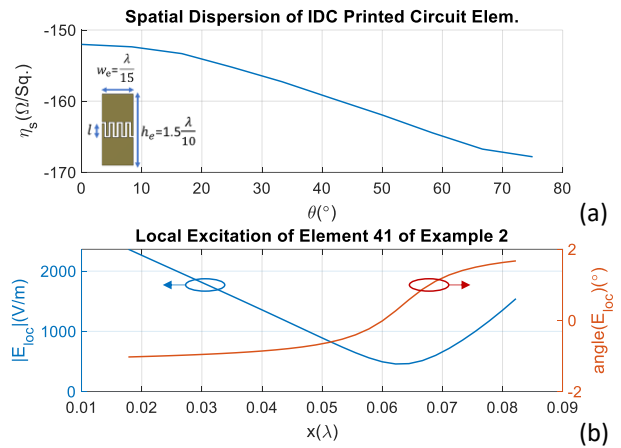

Fig. 1. Spatial dispersion of patterned metallic claddings and inhomogeneous excitation in aperiodic metasurfaces. (a) The spatial dispersion (dependence of extracted sheet impedance on tangential wavenumber of excitation) of interdigitated capacitor (IDC) with teeth length $l = 200\mu$m. The sheet impedance was obtained for an IDC in a locally periodic environment with unit cell dimension $\lambda/10 \times 1.5\lambda/10$ under a TE Floquet plane wave excitation incident at angle $\theta$. (b) The true inhomogeneous excitation of element number 41 of the metasurface example in section V due to the aperiodic neighbors.

Next-generation antennas designed using metasurfaces promise extreme wavefront control. Extreme wavefront control requires abrupt changes in field over subwavelength distances. Abrupt changes in field break the fundamental assumptions inherent to most commonly used extraction methods. Typically, fields are assumed to vary slowly, thereby allowing local periodicity approximations to be made. To realize the promise of metamaterials and metasurfaces, new design approaches which can account for rapid spatial field variations in design and in surface parameter extraction must be devised. Otherwise metasurface technologies will be forever relegated to the reduced dimensional space.

Metasurfaces utilizing printed circuit elements are typically designed in the homogenized/reduced-order domain. Subsequently, the homogenized elements are translated to patterned metallic claddings through a sheet impedance extraction procedure. The extractions are traditionally done in a locally periodic environment [5], [6]. This approach is strictly accurate only for periodic structures, approximate for adiabatic (slowly varying) structures, and fails completely when the metasurface is aperiodic with abrupt changes in field. The

Jordan Budhu is with the Department of Electrical and Computer Engineering, Virginia Tech, Blacksburg, VA, USA. Nicholas Ventresca, and Anthony Grbic are with the Department of Electrical Engineering and Computer Science, University of Michigan, Ann Arbor, MI, USA (e-mail: jbudhu@vt.edu, nickvent@umich.edu, agrbic@umich.edu).

difficulty stems from the aperiodic environment, which inhomogeneously excites the unit cells of the metasurface (see Fig. 1b). Fig. 1b shows the local electric field exciting a unit cell close to the center of an aperiodic metasurface reflectarray (see Section V). As is evident, the excitation cannot be described by a single plane wave. Since the printed-circuit geometries used to realize aperiodic metasurfaces are typically spatially dispersive (see Fig. 1a), this inhomogeneous excitation makes sheet impedance extraction challenging. When a metasurface's impedance profiles are found through optimization, as in [7]–[14], the impedance profile can exhibit abrupt changes (large relative impedance jumps between neighboring unit cells) that are needed to excite surface waves. These surface waves are responsible for redistributing power along the metasurface plane rendering it both passive and lossless [15]. Hence, the metasurface environment becomes highly aperiodic. New approaches or unit cell design procedures for these cases, which also inherently account for the printed circuit geometries spatial dispersion, must be developed to realize practical designs. This new approach will form a one-to-one relation between the reduced dimensional space and a realized prototype. Thus, it unlocks the promise of metamaterials and metasurface technology.

In this paper, an approach is presented for metasurface unit cell design in aperiodic environments. The approach begins with a solved linear system representing the homogenized electromagnetics problem (for example from the method of moments, finite element method, discrete dipole approximation, etc.). Each row of the solved linear system equates the total impedance (sheet plus self impedance) to the ratio of the local electric field and the induced surface current density. The inhomogeneous local field obtained from a row of the linear system is applied to a single parameterized, printed-circuit element as an impressed excitation. The induced surface current density is then calculated using full wave simulation. The surface current density is subsequently exported and sampled at the discretization of the linear system. The sampled surface current is inserted into the linear system in place of the homogenized surface current of the unit cell. Using the columns of the linear system, the field scattered by the printed circuit unit cell onto all neighboring unit cells and support structures (ground planes, dielectrics, etc.) is found. It is compared to that scattered from the homogenized unit cell. The geometrical parameter describing the printed circuit unit cell is varied until agreement between these scattered fields is made. In this way, a printed circuit element can be found which is a drop-in replacement for the homogenized element used in design.

Section II reviews current approaches to unit cell design in aperiodic metasurfaces. The section shows that the specific problem tackled in this paper is unsolved. In section III, the proposed approach is presented. In section IV and V, two aperiodic metasurfaces are realized using the proposed unit cell design approach. The performance is compared against the same metasurfaces realized using locally periodic material parameter extraction techniques. The paper concludes with section VI. An $e^{j\omega t}$ time convention is assumed and suppressed throughout the paper.

## II. CURRENT APPROACHES TO UNIT CELL DESIGN FOR APERIODIC METASURFACES

Unit cell design in aperiodic metasurfaces using canonical printed-circuit elements is described in [16]. Analytical formulas for the extracted sheet impedances of strips, patches, loops, and meandered lines are provided. The formulas are tested against periodic full wave simulation-based extraction techniques of the same geometries. The formulas were shown to be inaccurate for all but the strip element. Full wave optimization of the printed circuit dimensions was suggested as a remedy, and using the analytic formulas as a starting point. The authors do mention that the formulas are inaccurate for aperiodic metasurfaces, and one should ensure the sheet impedances vary adiabatically across the metasurface to reduce the error. Hence, their technique cannot handle general aperiodic metasurface unit cell design.

An approach to aperiodic metasurface unit cell design involving a combination of Particle Swarm Optimization (PSO) and Machine Learning (ML) is presented in [17]. Surrogate models resulting from the training data of several three-layer unit cell designs are made using a method of moments based electromagnetic solver. The surrogate models allow a Deep Neural Network (DNN) to predict the required unit cell geometry given the constituent material parameters the cell is intended to emulate. The predicted unit cell geometry is then fine-tuned using PSO to obtain the final unit cell geometry. The authors validate their unit cell design approach through experiment in [18]. Notable discrepancies arise between the far-field performance of the homogenized model and the metasurface made from the realized unit cells since the aperiodic environment is not accounted for in the realization step. Thus, this approach only approximately adheres to the performance of the homogenized models used in design. In some cases, the realized prototype failed to meet the targeted far-field objectives. The authors acknowledge that the lack of accurate mutual coupling modeling during the realization step is likely the cause of discrepancies. Hence, this technique also cannot handle general aperiodic metasurface unit cell design.

In [19], the systematic design of Huygens type unit cells consisting of collocated, orthogonal electric and magnetic dipoles is presented along with an experimental demonstration of plane wave refraction. The unit cells are designed using periodic boundary conditions, and hence the unit cell geometry must vary adiabatically for inhomogeneous metasurfaces. The measured results show notable discrepancies when compared with simulations. The authors attribute this to poor coupling between the feed horn and lens illumination system. As this technique is applicable to adiabatically-varying metasurface unit-cell geometries, it also cannot handle general aperiodic metasurface unit cell design.

In cases where the required constituent surface parameters do vary adiabatically, unit cell design can be approached using the locally periodic approximation. Examples are the metasurface designs featuring embedded sources in [14], [20]. Both metasurface designs, which transform the field radiated by a substrate embedded strip line feed to aperture fields exhibiting uniform amplitude and phase, were designed using inverse design coupled with integral equation modeling. They were realized using sheet impedance extraction techniques based on local periodicity and exhibited good agreement in simulation with the homogenized metasurfaces performance. However, when the surface parameters vary non-adiabatically, these same design approaches and locally-periodic based surface parameter extraction techniques produce large errors in the scattered fields. Full-wave cladding optimization is required to correct the scattered fields as evidenced in [7], [10]. For metasurfaces containing a larger number of elements, these full-wave optimizations of the entire cladding may not be feasible.

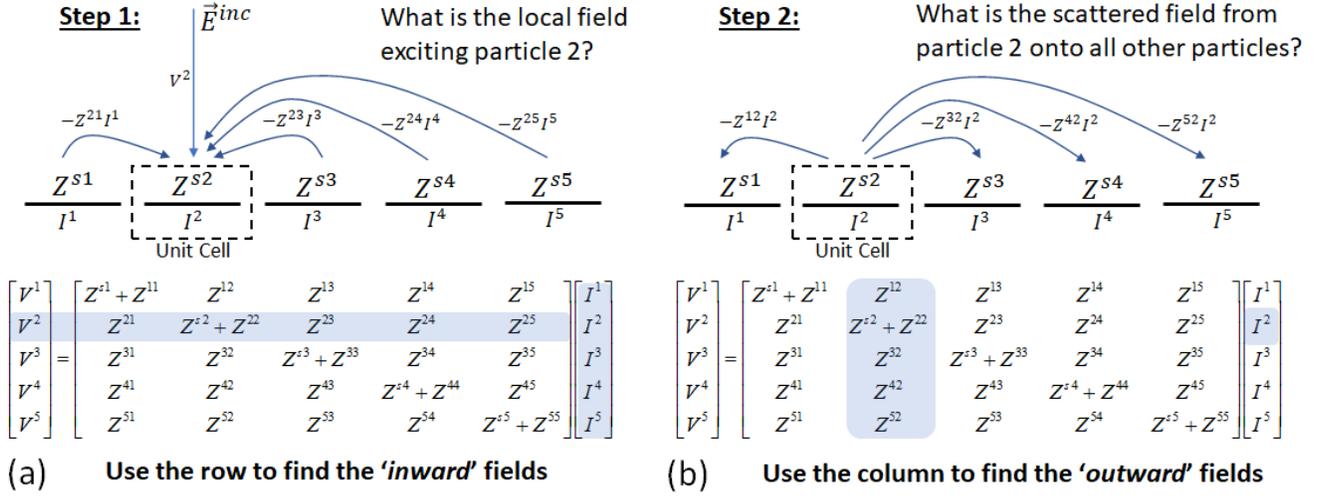

Fig. 2. Metasurface with 5 unit cells, and its matrix equation representation. (a) Step 1: Use the row of the solved block matrix equation to derive the local electric field exciting a unit cell. (b) Step 2: Use the column to determine the field scattered by a unit cell onto its neighbors. Each block is $K \times K$, where $K$ is the number of unknowns placed on each homogenized sheet. The shaded row/column is for the $p = 2$ element.

A metasurface is said to be aperiodic when either the neighboring unit cells on a regular grid is unique or when the elements are placed on an irregular grid. A technique presented in [21] allows for aperiodic metasurface design on irregular grids. By modeling the metasurface elements as discrete dipoles, the polarizability model can be used to compute the dipole moment given the exciting local electric field. The local electric field is computed by a combination of dipole summation for the nearest neighbors, and current sheet homogenization for the far elements [22]. The desired scattered fields are expressed in terms of the induced dipole moments. The metasurface can be designed, element-by-element, for any grid placement of the dipoles. This approach has the limitation of being applicable to metasurfaces made from elements which can be well modelled using the dipole-dipole interactions. This is not the case for densely-packed, subwavelength printed-circuit geometries comprising most metasurfaces in current literature. Hence, this technique also cannot handle general aperiodic metasurface unit cell design.

In this paper, we present an approach which models the surface currents on printed-circuit geometries without resorting to dipole approximations and is thus applicable to a wider range of designs. The metasurface is broken into equal sized and spaced unit cells, each of which contain a unique printed circuit element. The unit cells can vary non-adiabatically with large changes in neighboring cell geometries. The details of the proposed unit cell design approach are provided in the next section.

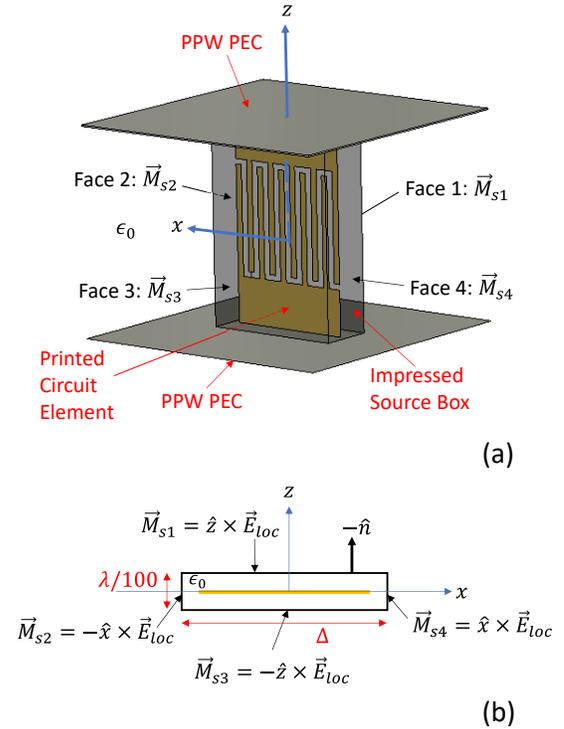

Fig. 3. Magnetic sheet current box for exciting a unit cell with the local field. (a) A 3D view of the printed circuit geometry within the magnetic sheet current box. (b) The cross sectional view of the same. The magnetic sheet current box is designed to bathe the printed circuit in the inhomogeneous local field within the box and produce zero field outside of the box.

## III. THE PROPOSED APPROACH TO UNIT CELL DESIGN FOR APERIODIC METASURFACES

The proposed approach to unit cell design is detailed in this section. The starting point is a metasurface design consisting of a spatially-variant, homogenized impedance sheet in the presence or absence of a grounded substrate. The metasurface is modeled by a matrix equation. In our case, the matrix equation is obtained from the method of moments (MoM), and solved for all unknowns, as in the designs of [13], [14]. For an overview of the design of aperiodic, metasurfaces using integral equations, see [23].

As an example, consider the metasurface illustrated in Fig. 2. The figure shows a 1D array of 5 homogenized sheet impedances. The electromagnetic problem described here is 2D, and hence the out-of-plane wavenumber is equal to 0. However, the approach is applicable to 3D problems as well. As depicted in Fig. 2, Step 1 uses the rows of the matrix equation to find the local field exciting each unit cell. This local field is then impressed onto a printed-circuit unit cell in a separate full wave simulation. The surface current density is exported from the full-wave simulation and imported (sampled at the MoM discretization) into the current coefficient vector on the right hand side of the matrix equation. Step 2 uses the columns of the matrix equation to calculate the field scattered from the printed-circuit unit cell onto its neighboring elements. This calculation is used to form a cost function comparing the scattered fields from the homogenized unit cell and the printed-circuit unit cell. The cost function is used in a 1-parameter optimization to determine the unit cell geometry which couples to all neighboring elements equivalently to the homogenized sheet. Since the optimization is over only one parameter (the geometrical variable describing the printed circuit parameterization), the optimization is fast and converges in only a few iterations. Since the full-wave simulation is performed over a single unit cell, each iteration is computed rapidly, making the entire realization (patterning) process practical even for large metasurfaces. The details of the approach are described next.

### *Step 1 Details:*

Consider a block matrix equation similar to that shown in Fig. 2 that models the electromagnetic problem of an excited, homogenized metasurface. Each block is $K \times K$, where $K$ is the number of unknowns placed in each homogenized unit cell. Consider the equality constructed from the $p^{th}$ row of the block matrix equation. It describes the response of the $p^{th}$ element of the metasurface. In Fig. 2, this equality is highlighted in blue for the $p = 2$ element. The total impedance (sheet impedance plus the self impedance) of element $p$ can be obtained by moving all terms except the self-scattering term ($p = q$) to the left-hand side of the equation, and dividing (element-wise) by the induced current density on element $p$, namely,

$$\frac{E_{loc}^p}{J^p} = \frac{V^p - \sum_{q \neq p} Z^{pq} I^q}{I^p} = \left( Z^{pp} + Z^{sp} \right) = \eta_{aperiodic}(l) \quad (1)$$

The numerator of the left hand side is the inhomogeneous local electric field. For example, this is how the plot in Fig. 1b was calculated. The local electric field, $E_{loc}^p$, is impressed onto a parameterized (by $l$) printed-circuit geometry using the magnetic current sheet box depicted in Fig. 3, and the surface current density induced on the printed circuit geometry is computed using a full-wave solver, in this case COMSOL Multiphysics. Since the problem is 2D, the printed circuit can be placed in a parallel plate waveguide. To incorporate the effects of the backing ground plane, an image printed circuit geometry is placed at $y = -2d$ and excited with $-E_{loc}$. The computed surface current density is then exported as a function of both the $x$ and $z$ coordinates. The exported surface current density, $J_s^{COMSOL}(x,z)$, is then homogenized (averaged) vertically and sampled at the moment method (matrix equation) discretization

$$I^{p,\text{PrintCir}} = \left[ I_1^{p,\text{PrintCir}}, I_2^{p,\text{PrintCir}}, \ldots, I_k^{p,\text{PrintCir}}, \ldots, I_K^{p,\text{PrintCir}} \right]^T$$

$$= \int_{-w_e/2}^{w_e/2} \delta \left( x + \frac{w_e}{2} - \frac{k\Delta x}{2} \right) \left[ \frac{1}{h_e} \int_{-h_e/2}^{h_e/2} J_s^{COMSOL}(x,z) dz \right] dx \quad (2)$$

where $k=1,2,\ldots,k,\ldots,K$ is a sample index, $K$ is the total number of unknowns placed horizontally on the $p^{th}$ homogenized sheet in the moment method problem, $\Delta x$ is their separation, and $h_e$ and $w_e$ are the printed circuit geometry width and height (see inset of Fig. 7 or Fig. 12). Application of (2) makes the current density of the printed circuit geometry compatible with the matrix equation modeling the 2D electromagnetic problem. It can then be substituted in for $I^p$ in the current coefficient vector on the right hand side of matrix equation. In Fig. 2b, this is shown as the blue shaded block $I^2$ for $p = 2$ in the current coefficient vector.

### *Step 2 Details:*

The $p^{th}$ column of the block matrix equation is then used to calculate the scattered field, $E_{scat}^p$, due to the current density in (2) onto its neighboring elements and ground plane. A cost function is formed which compares the fields scattered by the homogenized unit cell (impedance sheet) to the printed-circuit unit cell

$$f = \left( \Delta E_{scat}^p \right)^\dagger \left( \Delta E_{scat}^p \right) + \left( \Delta E_{nf}^p \right)^\dagger \left( \Delta E_{nf}^p \right) \quad (3)$$

where † indicates Hermitian transpose and

$$\Delta E_{scat}^p = \sum_{q \neq p} Z^{qp} \left( I^{p,\text{ImpSheet}} - I^{p,\text{PrintCir}} \right) \quad (4)$$

Note, $I^{p,ImpSheet}$ is obtained directly from the solved matrix equation describing the homogenized metasurface. Since (4) does not consider self scattering, the second term of (3) compares the near fields just above element $p$, namely, $\Delta E_{nf}^p = E_{nf}^{p,ImpSheet} - E_{nf}^{p,PrintCir}$. The near fields are analytically calculated from the current densities $I^{p,ImpSheet}$ and $I^{p,PrintCir}$ using the 2D Green's function, along a horizontal line $\lambda/2$ wide and $\lambda/5$ above the element. The cost function (3) is minimized as a function of the length of the interdigitated metallic fingers, $l$, of the interdigitated capacitor, for example. This process is repeated for each element $p$ in the metasurface to obtain the patterned metallic cladding.

## IV. EXAMPLE 1: A WIDE-ANGLE REFLECTING METASURFACE

An aperiodic metasurface with fast field variation is a finite width, wide-angle reflecting metasurface. The metasurface considered is shown in Fig. 5. The metasurface consists of 54 unit cells each of width $1.5\lambda/10$ along the $x$-direction. Each unit cell is filled with an IDC of width $w_e = \lambda/15$. The unit cells are $h_e = 1.5\lambda/10$ tall in the $z$-direction and periodic along that direction. The metasurface is separated from the $7.8\lambda$ wide ground plane by a $\lambda/20$ thick Rohacell HF31 foam substrate with relative permittivity $\epsilon_r = 1.04$ (approximated as free space in design). A normally incident plane wave at 10 GHz is reflected to an angle of 70°. The design of this metasurface is challenging since a plane wave illumination strongly illuminates the finite edges of the metasurface and produces strong diffraction which must be cancelled by the metasurface. The metasurface is designed using the three-phase design approach outlined in [14], [23]. In summary, the three-phase design approach involves (a) a direct-solve solution of an integral equation modeling the electromagnetics problem. It often results in a metasurface requiring local loss and/or gain [24], (b) a subsequent optimization phase to convert the metasurface designed in phase 1 to a purely passive and lossless design, and (c) a metallic patterning phase (the subject of this paper).

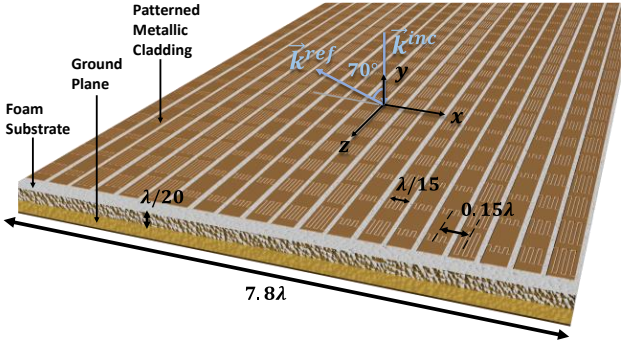

Fig. 5. Representative figure of a finite-width, wide-angle reflecting metasurface of example 1. The geometry is infinite and invariant in the $z$-direction and finite and spatially variant in the $x$- and $y$-directions. The metasurface contains 3 layers, one patterned metallic cladding layer suspended above a grounded Rohacell foam substrate with near unity relative permittivity. (Note: metasurface elements have been enlarged for clarity and visibility and therefore fewer are shown than the actual metasurface. The actual metasurface contains fifty-four $\lambda/15$ wide elements, each separated by $0.15\lambda$).

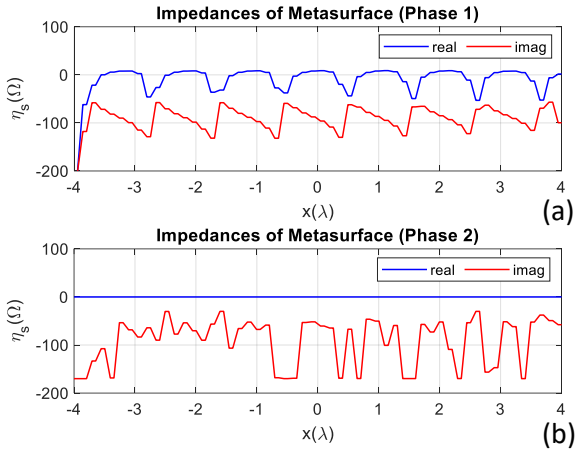

Fig. 6. Metasurface sheet impedances. (a) Complex sheet impedance of Phase 1 obtained by directly solving the governing matrix equation. (b) Optimized reactive sheet impedance of Phase 2.

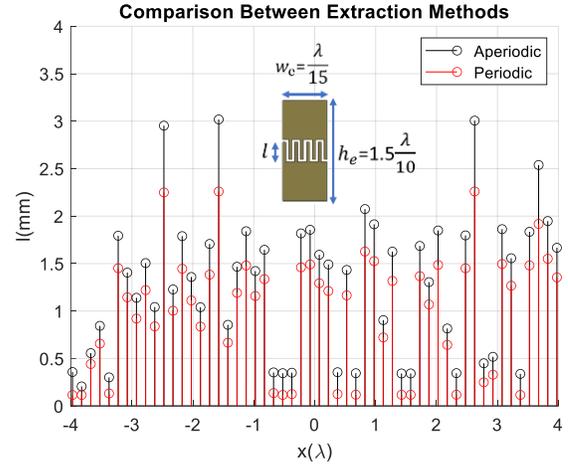

Fig. 7. Comparison between the claddings obtained from locally periodic extractions techniques and the proposed aperiodic unit cell design technique.

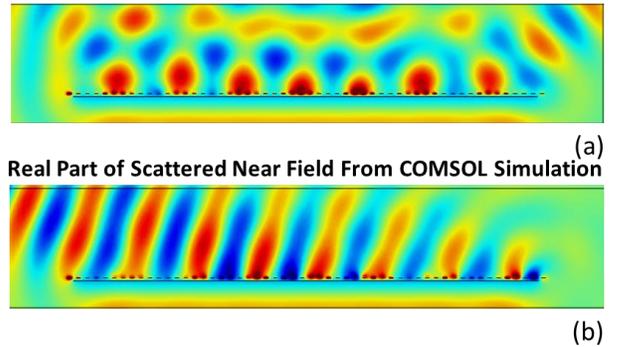

Fig. 8. Real part of the scattered near field for the metasurface constructed from the optimized reactive metasurface sheet impedances of Fig 6b. (a) metasurface realized using the sheet impedance extraction unit cell design technique based on local periodicity. (b) metasurface realized using the proposed aperiodic unit cell design technique.

The direct solve solution of the integral equation in Phase 1 leads to the complex-valued sheet impedance shown in Fig. 6a. Gradient descent optimization, accelerated with the adjoint variable method, is applied in Phase 2 to convert the complex-valued sheet into a purely reactive sheet. For details on the optimization procedure, see [14]. The optimization introduces surfaces waves by perturbing the reactances (the initial resistances are discarded) of the metasurface until a purely reactive boundary condition is satisfied that generates the same far field as the complex-valued sheet from the direct solve solution [10], [12]–[15]. The result is a metasurface with a highly aperiodic sheet reactance profile, as shown in Fig. 6b. It performs identically to the complex-valued sheet impedances of Fig. 6a at the distances of interest. One advantage of the optimization is that bounds can be placed on the sheet reactances to align with realizable values. This way the optimized reactive metasurface is guaranteed to be realizable through patterning of metallic claddings. In the example, the reactances were constrained to the interval $-j170\Omega \leq \eta_s \leq -j30\Omega$.

The unit cell design procedure was used to obtain the patterned metallic cladding for the reactive sheet impedance profile shown in Fig. 6b. The patterned metallic cladding is shown compared to the realization using traditional locally periodic extraction based approaches in Fig. 7. As can be seen, the unit cell design approach produces a different pattern than locally periodic based approach which does not consider the spatial dispersion inherent to the printed circuit geometry due to the inhomogeneous excitation.

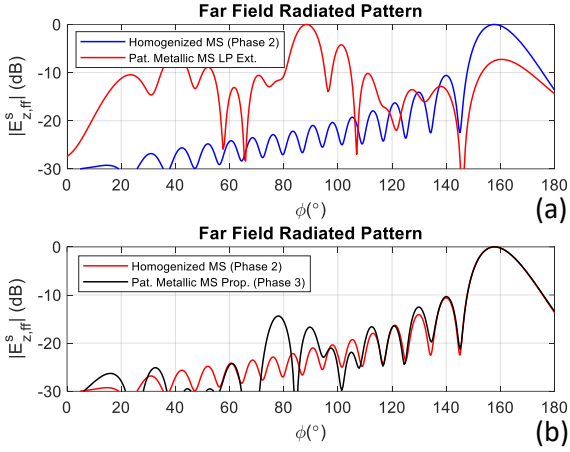

Fig. 9. Far field pattern of the optimized reactive impedance sheet design compared to (a) the COMSOL Multiphysics full-wave simulation of the patterned metallic cladding obtained from locally periodic extractions, and (b) the COMSOL Multiphysics full-wave simulation of the patterned metallic cladding obtained from the proposed aperiodic unit cell design.

Both the metasurfaces designed using the locally periodic and aperiodic unit cell design approaches were simulated in COMSOL Multiphysics. The resulting real part of the scattered near field is shown in Fig. 8 for both cases. As can be seen, the metasurface designed using locally periodic based extractions (Fig. 8a) fails to perform wide-angle reflection, whereas the metasurface designed using the aperiodic unit cell design technique (Fig. 8b) does. This is further evident in Fig. 9, which shows the scattered far field superimposed over the far field of the metasurface containing the homogenized impedance sheet. Again, we see the unit cell extraction technique based on local periodicity [5], [6] does not produce the desired far field, whereas the aperiodic unit cell extraction technique does.

## V. EXAMPLE 2: A LINE SOURCE FED REFLECTING AND COLLIMATING METASURFACE

As another example, consider the metasurface shown in Fig. 10. This metasurface contains 80 unit cells each $\lambda/10$ wide in the $x$-direction. Each unit cell contains an IDC of width $\lambda/15$. The metasurface is supported $\lambda/20$ above a perfectly conducting ground plane by a Rohacell HF31 foam substrate. The ground plane is $8\lambda$ wide. The metasurface is fed by a 10 GHz electric line source placed along the $y$-axis at $F = 2\lambda$ above the metasurface. The metasurface is designed using the same three-phase design strategy to transform the incident cylindrical wave into a reflected plane wave traveling at an angle of $30°$ with respect to the metasurface normal ($\hat{y}$). The complex-valued impedance sheet of Phase 1 is shown in Fig. 11a, and the optimized reactive impedance sheet result of Phase 2 is shown in Fig. 11b. The comparison of the cladding designed using the proposed aperiodic unit cell design approach to that designed using local periodicity is provided in Fig. 12. As can be seen, the claddings are different. The far field resulting from both claddings are compared in Fig. 13. The figure shows the unit cell design approach utilizing local periodicity fails to recreate the sidelobes of the homogenized design, whereas the proposed unit cell design technique does. Again, without the proposed unit cell design technique, finite aperiodic metasurfaces cannot be realized with close adherence to the homogenized models used in design.

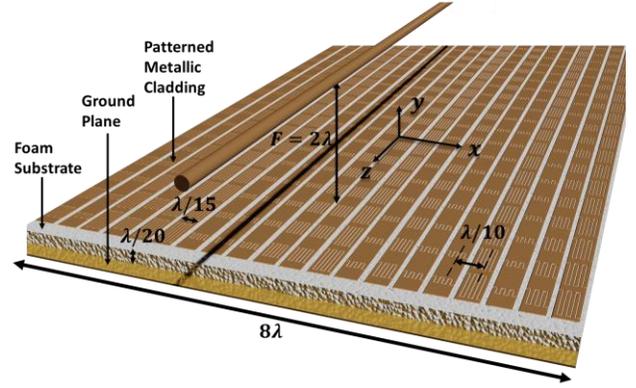

Fig. 10. Representative figure of the metasurface reflectarray of example 2. The geometry is infinite and invariant in the $z$-direction and finite and spatially variant in the $x$- and $y$-directions. The metasurface contains 3 layers, one patterned metallic cladding layer suspended above a grounded Rohacell foam substrate with near unity relative permittivity. (Note: metasurface elements have been enlarged for clarity and visibility and therefore fewer are shown than the actual metasurface. The actual metasurface contains eighty $\lambda/15$ wide elements, each separated by $\lambda/10$.)

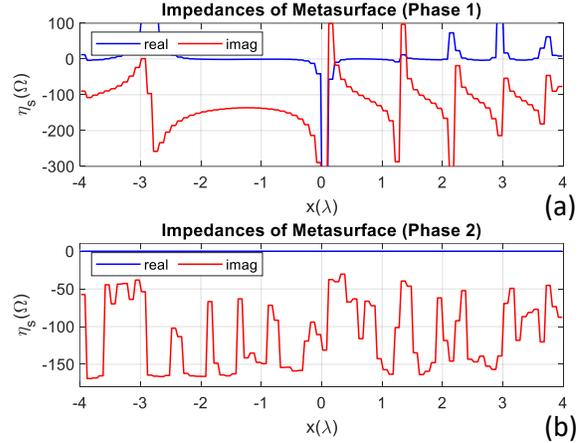

Fig. 11. Metasurface sheet impedances. (a) Complex sheet impedance of Phase 1 obtained by directly solving the governing matrix equation. (b) Optimized reactive sheet impedance of Phase 2.

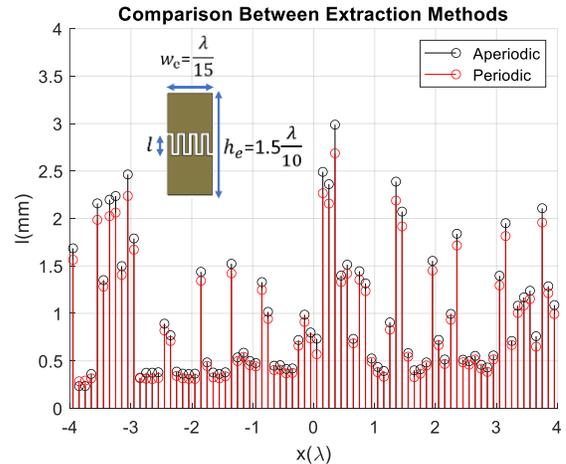

Fig. 12. Comparison between the claddings obtained using the locally periodic extraction technique and the proposed aperiodic unit cell design technique.

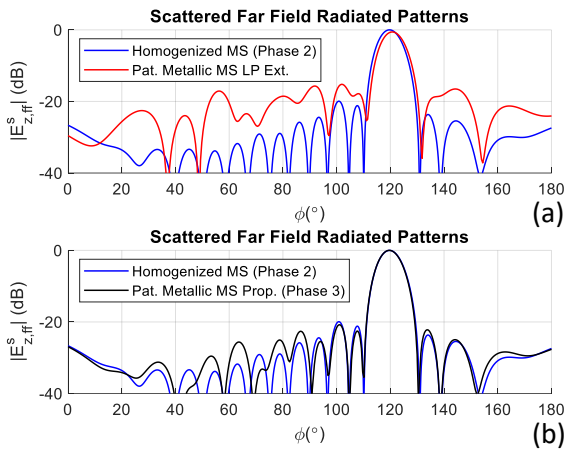

Fig. 13. Far field pattern of the optimized reactive impedance sheet design compared to (a) the COMSOL Multiphysics full-wave simulation of the patterned metallic cladding obtained from locally periodic extractions, and (b) the COMSOL Multiphysics full-wave simulation of the patterned metallic cladding obtained from the proposed aperiodic unit cell design.

## VI. CONCLUSION

A new approach to designing printed-circuit unit cells in highly aperiodic metasurfaces was presented. The approach starts with a solved linear system that models the homogenized metasurface. The inhomogeneous, local, electric field exciting an individual element in the aperiodic metasurface environment is found using the matrix equation. This local field is used as the impressed excitation in a separate full-wave simulation to compute the induced surface current on the printed circuit element. The surface current density is sampled at the discretization of the matrix equation and exported. The matrix equation is then used to find the scattered field onto all neighboring elements. The printed circuit element of each unit cell is optimized to scatter the same fields onto its neighboring unit cells as the homogenized metasurface. Two metasurfaces realized using the proposed unit cell design approach show excellent agreement with the homogenized sheet model. It was shown that the same agreement cannot be obtained using conventional unit cell design techniques based on local periodicity.